\documentclass[conference]{IEEEtran}
\IEEEoverridecommandlockouts
\usepackage{amsmath,amssymb,amsfonts}
\usepackage{pifont}
\usepackage{blindtext}
\newcommand{\cmark}{\ding{51}}
\newcommand{\xmark}{\ding{55}}
\usepackage[section]{placeins}
\usepackage{booktabs}
\usepackage{longtable}
\usepackage[utf8]{inputenc} 
\usepackage{graphicx} 
\usepackage{xcolor}
\usepackage[ruled,linesnumbered]{algorithm2e}
\SetKwProg{Init}{Initialization:}{}{}
\graphicspath{{figures/}} 
\usepackage{verbatim}

\begin{document}

\title{MIX-MAB: Reinforcement Learning-based Resource Allocation Algorithm for LoRaWAN}

\author{Farzad Azizi$^*$, Benyamin Teymuri$^*$, Rojin Aslani$^*$, Mehdi Rasti$^*$, Jesse Tolvanen$^\dagger$, and Pedro H. J. Nardelli$^\dagger$
\\ $^*$Department of Computer Engineering, Amirkabir University of Technology, Tehran, Iran
\\ $^\dagger $School of Energy Systems, {Lappeenranta-Lahti University of Technology}, Lappeenranta, Finland 
\\ Email: {\{farzad.azizi, 
benyamin.teymuri, rojinaslani, rasti\}@aut.ac.ir, \{jesse.tolvanen, pedro.nardelli\}@lut.fi}
\thanks{This paper is partly supported by Academy of Finland via (a) FIREMAN consortium n.326270 as part of CHIST-ERA grant CHIST-ERA-17-BDSI-003, and (b) EnergyNet Research Fellowship n.321265/n.328869 and (c)  n.339541, and by Jane and Aatos Erkko Foundation via STREAM project.}} 
\maketitle

\begin{abstract}
This paper focuses on improving the resource allocation algorithm in terms of packet delivery ratio (PDR), i.e., the number of successfully received packets sent by end devices (EDs) in a long-range wide-area network (LoRaWAN). Setting the transmission parameters significantly affects the PDR. Employing reinforcement learning (RL), we propose a resource allocation algorithm that enables the EDs to configure their transmission parameters in a distributed manner. We model the resource allocation problem as a multi-armed bandit (MAB) and then address it by proposing a two-phase algorithm named MIX-MAB, which consists of the exponential weights for exploration and exploitation (EXP3) and successive elimination (SE) algorithms. We evaluate the MIX-MAB performance through simulation results and compare it with other existing approaches. Numerical results show that the proposed solution performs better than the existing schemes in terms of convergence time and PDR.
\end{abstract}
\begin{IEEEkeywords}
IoT, LPWAN, LoRaWAN, LoRa, distributed resource allocation, reinforcement learning, multi-armed-bandit.
\end{IEEEkeywords}
\IEEEpeerreviewmaketitle
% \vspace{-5mm}
\section{Introduction}
The maturity of internet of things (IoT) technology is already rapid. According to projections for the next ten years, over $125\times10^9$ IoT devices are expected to be connected worldwide \cite{nizetic}. The Low-power wide-area network (LPWAN) can provide the network connection for many end devices (EDs) in a wide range consuming low battery power \cite{Babaki}. Many protocols for LPWAN exist including long-range wide area network (LoRaWAN) \cite{Babaki}, SigFox \cite{Sigfox}, and NB-IoT. LoRaWAN is one of the most promising candidates for IoT, attracting more~attention~and~can~support~many~applications~such~as~electricity~metering, localization, and industrial applications [4-5]. \par The transmission parameters, i.e., radio resources, including spreading factors (SFs), sub-channels (SCs), and transmission power (TP), have a significant role in determining the throughput of LoRaWAN \cite{performance}. Legacy LoRa runs an adaptive data rate (ADR) mechanism in the central controller, i.e., a network server (NS), to compute the transmission parameters of EDs and send them back as a MAC command \cite{LoRaWAN}. However, frequent communication between EDs and the NS is inefficient because of the EDs' limited energy and duty cycle (DC) restrictions. Furthermore, the NS cannot respond to all the messages due to strict deadlines. Therefore, it is essential to design a distributed resource allocation algorithm in LoRaWAN. \par Machine learning makes it possible to use distributed algorithms. Reinforcement learning (RL) is a machine learning technique that does not need any training data sets, making it the best choice for LoRaWAN. The literature review shows that RL techniques can improve resource allocation performance in LoRaWAN by enabling each LoRa ED to select the most suitable configuration settings through the self-learning process. More especially in [8-9], a non-stationary RL-based resource allocation algorithm called LoRa-MAB is proposed, using an adversarial environment similar to LoRaWAN's circumstances. However, LoRa-MAB suffers from high convergence time due to the long exploration process. This paper shows that combining the non-stationary adversarial algorithms, suitable for LoRaWAN environment, with stochastic ones, having the advantage of a short exploration process, improves the convergence time and increases the packet delivery ratio (PDR) in LoRaWAN. Contributions of this paper are as follow:
% \vspace{-5mm}
\begin{itemize}
    \item We model the resource allocation problem in LoRaWAN as a Multi-armed bandit (MAB) problem. Then we employ a mixture of exponential weights for exploration and exploitation (EXP3) as a non-stationary adversarial scheme and the successive elimination (SE) as a non-stationary stochastic scheme to propose our solution called MIX-MAB.
    \item Through simulation results, we compare the performance of our algorithm with LoRa-MAB \cite{Ta} and Legacy LoRa \cite{Bor} approaches in terms of convergence time, energy consumption (EC), and PDR. Simulation results show that the convergence time of the MIX-MAB is half of the LoRa-MAB while achieving the higher PDR than others in different scenarios. 
    \item We also present a simulation scenario in which EDs have the freedom to select their SFs, SCs, and TP simultaneously. Our numerical results show that the proposed MIX-MAB algorithm achieves high PDR with the lowest EC in such a scenario compared to those where the EDs can configure only one or two transmission parameters.
    \item Finally, the effect of the number of packets sent by each ED is examined, showing the more frequently transmitted packets leads to reduced the PDR and the EC.
\end{itemize}
\begin{figure}[t]
    \begin{center}
        \includegraphics[width=0.45\textwidth]{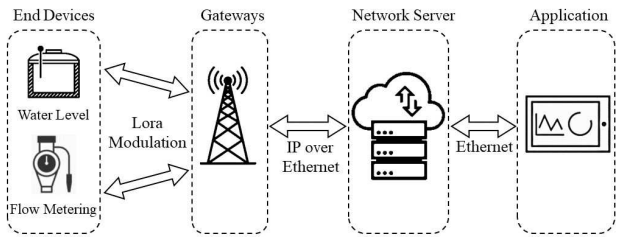}
    \end{center}
    \vspace{-5mm}
    \caption{LoRaWAN network architecture.}
    \label{fig:arc}
\end{figure} 
\setlength{\textfloatsep}{0pt}
% \vspace{-5mm}
\par Rest of this paper is organized as follows. Section II presents the background and related works. Our proposed MIX-MAB algorithm is described in Section III. Sections IV and V contain the numerical results and conclusion, respectively. 
% \vspace{-5mm}
\section{Background and Related Works}
This section reviews LoRa physical layer and transmission parameters in LoRaWAN. Then, it examines the related works. 
% \vspace{-5mm}
\subsection{LoRa and LoRaWAN overview}
LoRaWAN is a long-range wireless interconnections system composed of two core components: LoRa and LoRaWAN. LoRa is a physical layer using the chirp spread spectrum (CSS) radio modulation developed by Semtech \cite{Sigfox}. LoRaWAN, implemented on top of the LoRa, includes data link and network layers using star network topology for data transmission. LoRaWAN network architecture is shown in Fig.\ref{fig:arc}. As seen, the EDs send data to the NS through the gateways (GWs). LoRaWAN~specifications~are~documented~by~LoRa~Alliance~\cite{LoRaWAN}. 
% \vspace{-5mm}
\subsection{LoRa transmission parameters} 
In LoRa physical layer, each transmission depends on the following parameters:
\begin{itemize}
    \item \textit {\textbf{SFs:}} 
    EDs can select an SF value from 7-12 based on the environmental condition between the ED and the GW. Selecting SF creates a trade-off between the data rate, communication range, and energy utilization.
    \item \textit {\textbf{SCs:}} Depending on the world region, LoRa communications can operate over license-free sub-GHz radio frequency bands including $\{433,~868,~915\}$ MHz.
    \item \textit {\textbf{TP:}}
    The LoRa radio TP is adjustable from $-4$ to $20$ dBm in steps of $1$ dBm.
\end{itemize} 
Successful transmissions in LoRaWAN, leading to the raised PDR, rely on the interference management performed by adequately adjusting the transmission parameters depending on the conditions between the EDs and GWs. There are two ways to control transmission parameters in LoRaWAN: link-based approach and network-aware approach \cite{Babaki}. In the link-based approach, transmission parameters are asynchronously configured in a centralized manner by commands that ED receives from the NS \cite{Li}. In the network-aware method, each ED configures its transmission parameters in a distributed way.
% \vspace{-5mm}
\subsection{Related Works}
Several research efforts have improved LoRa/LoRaWAN performance, focusing on optimization or performance analysis [7-11]. The ADR algorithm has been proposed in \cite{LoRaWAN} for setting the LoRa transmission parameters. The proposed approaches in \cite{Babaki} and \cite{Li} worked on improving the original ADR algorithm by using the history of received packets in the NS. However, the ADR mechanism as a centralized link-based approach has two shortcomings. The first one is that its PDR decreases by increasing the number of EDs. The second one is the increased EC in dens deployment of IoT EDs. Current research has focused on using machine learning techniques such as RL, enabling EDs to use innovative and inherently distributed techniques, thus preventing them from evacuating their limited power by permanently communicating with the NS [8-9], \cite{Aihara}. The MAB [8-9] and Q-learning \cite{Aihara} are two RL algorithms used in the literature to propose distributed radio resource allocation in LoRaWAN. In \cite{Aihara}, authors applied Q-learning to offer a resource allocation for LoRaWAN, aiming at decreasing the collision rate and improving the network PDR. However, despite increasing the EC using the method in \cite{Aihara}, the PDR is still a function of available channels. Moreover, the Q-Learning solution requires the database to save its processing data. The LoRa-MAB algorithm proposed in \cite{LoRa-MAB} suffers from a severe drawback of high convergence time (equal to $200$ kHours as discussed in \cite{Ta}). The SE is a MAB-based algorithm presented in \cite{Allesiardo} while the adversarial environment of LoRa has not been taken into account.
\section{Our proposed MIX-MAB algorithm}
This section proposes a distributed resource allocation algorithm in LoRa. Machine learning techniques as a distributed approach are divided into supervised, unsupervised, and reinforcement learning. There is no need to train data sets in RL-based methods, where learning happens through interaction with the environment. RL agent can perceive and analyze its~environment,~take~actions~and~learn~through~trial~and~error \cite{Sutton}. We employ RL to propose our resource allocation algorithm. Therefore, there is no need to provide predefined data to EDs, and they will learn through sending messages on the network. Thus, our proposed algorithm imposes no computational overhead on~the~network~from~this~point~of~view. \par In LoRaWAN, the agents are the LoRa EDs interacting with the environment, including GWs and the other EDs, to take actions defined as selecting their transmission parameters set. The agent learns the best actions based on the received reward defined based on the acknowledgment (ACK) messages. More specifically, a LoRa ED selects a set of transmission parameters and sends its data packet based on the selected setting. If the NS receives the packet, it sends a confirmation ACK message to the ED. Receiving the ACK, the ED assigns a binary reward (defined later) to the selected action set and uses it for subsequent transmission parameters index. \par We model the transmission parameters configuration by LoRa EDs as a MAB problem, an RL-based technique, and formulated it using k-armed bandits. Accordingly, an agent selects from k-different actions and receives a reward based on its chosen action.\par Three general categories of stochastic, adversarial, and switching bandit algorithms exists to address the MAB problems. EXP3 is a non-stationary adversarial MAB problem. LoRa EDs are placed in this type of algorithm because selecting the same parameter like SFs affects the other EDs. But the lack of a short exploration process in this approach will expand the convergence time. Stochastic algorithms such as SE are not suitable for the LoRaWAN individually because of their adversarial nature. However, the long exploration process of EXP3 results in a high convergence time, which is a critical performance criterion in LoRaWAN. On the other hand, the SE algorithm has the advantage of short convergence time due to a short exploration process. So, Inspired by the benefits of EXP3 and SE algorithms used in \cite{Ta} \cite{Allesiardo}, respectively, we combine these two approaches and propose a new algorithm called MIX-MAB. \par We assume that there are $U$ EDs in the LoRaWAN network forming the set of $\mathcal{U}=\{1,2,\cdots,U\}$. Each ED aims at maximizing its PDR in a decentralized manner by learning to select the most appropriate transmission parameters set. \par Let $\mathcal{S}$, $\mathcal{C}$, and $\mathcal{P}$ denote the set of SFs, SCs, and TP, respectively. Assuming that each action is a vector composed of three parameters representing SF, SC, and TP, $I_k=\{s_k,c_k,p_k\}$ denote the $k$th vector of parameters, means $k$th action, in which $s_k \in \mathcal{S}$, $c_k \in \mathcal{C}$, and $p_k \in \mathcal{P}$ are the values of SF, SC and TP in $k$th action, respectively. We assume that there are $K$ actions whose set denoted by $\mathcal{K}=\{I_0, I_1, \cdots, I_{K-2}, I_{K-1}\}$. Let $A_u(t) \in \mathcal{K}$ describe the selected action of the $u$th ED at the $t$th iteration. Each iteration corresponds to a packet arrival in the ED. Taking the $k$th action, the $u$th ED receives the related reward presented by $R^u_k(t) \in \{0,1\}$, where $R^u_k(t)=1$ for successful transmissions, i.e., the ACK message is received, while receiving the NACK message is denoted by $R^u_k(t)=0$ for fail transmissions. We assign a probability of selection to each action. Let $P^u_k(t)$ be the probability of taking the $k$th action by the $u$th ED at the $t$th iteration. To obtain the probabilities, we assign a weight to each action. $W^u_k(t)$ is the weight of the $k$th action selected by the $u$th ED at the $t$th iteration. The goal of each ED $u$ is to update $P^u_k(t)$ to achieve the largest reward during $T$ iterations. \par
\begin{figure}[t]
    \centering
    \includegraphics[scale=0.35]{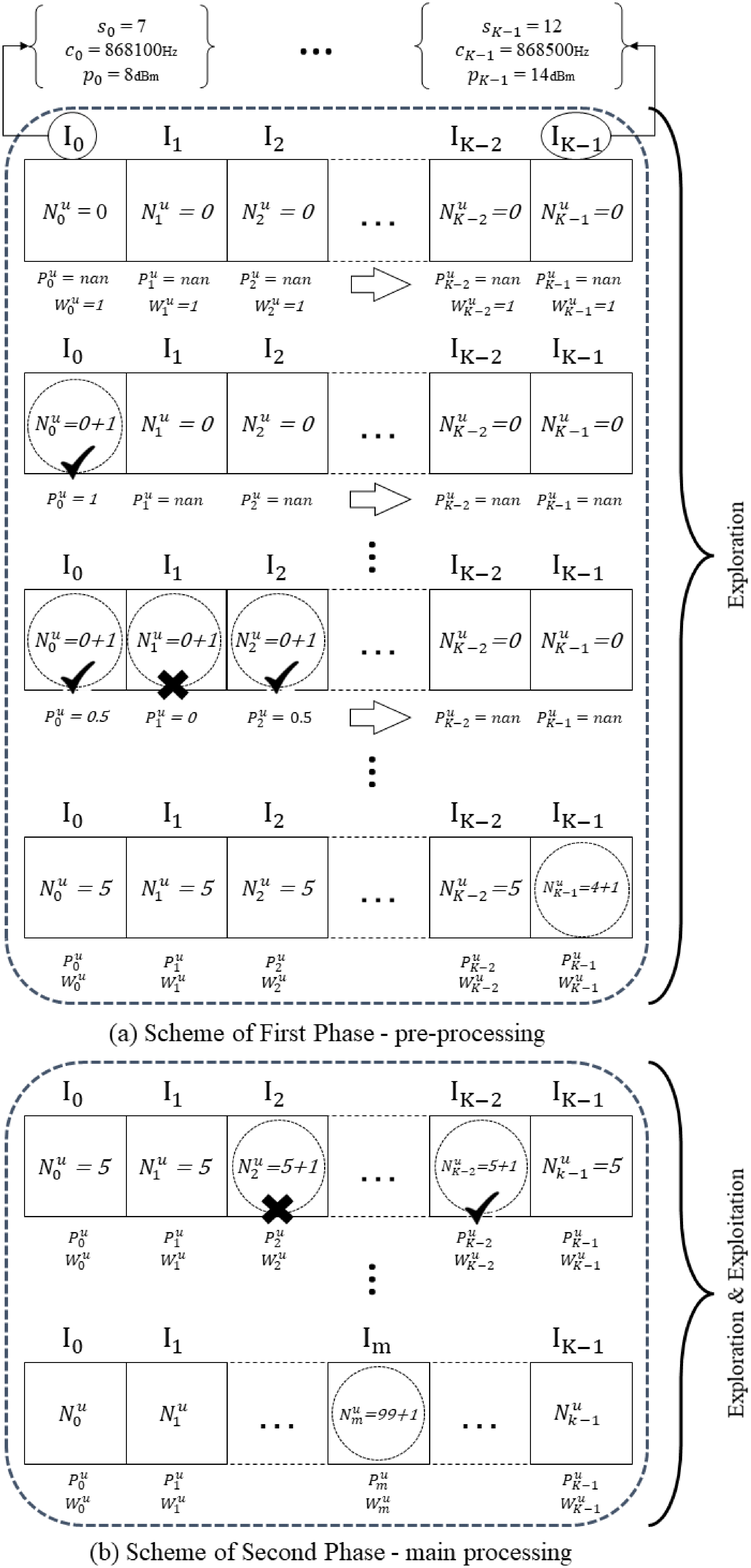}
    % \vspace{-8mm}
    \caption{Proposed two-phase solution scheme for the $u$th ED.}
    \label{fig:rl}
\end{figure}

 Our~proposed~MIX-MAB algorithm presented in Algorithm \ref{al}~composed~of~two~phases.~The~first phase,~called~the~pre-processing~phase,~only~includes~exploration process. The second phase, called the main processing phase, consists of exploration and exploitation processes. Fig. \ref{fig:rl} illustrates the proposed MIX-MAB scheme for the $u$th ED as explained in what follows.
\setlength{\textfloatsep}{0pt}
% \vspace{-5mm}
% \vspace{-5mm}
\begin{algorithm}[t]
    \caption{MIX-MAB}
    \label{al}
    \SetAlgoLined
    \Init{}{
    Set $u \in \mathcal{U}$ as a $u$th ED \\
    Set learning rate $\gamma \!=\! \min \Big\{1,\sqrt \frac{K\log(K)}{(e-1)T} \Big\}$, $e\!=\!2.71$ \\
    Set $l^\textnormal{EXP}=5, \alpha_u = 1, l^\textnormal{EE}=100$, \\
    Set $P^u_k(0)=nan, N^u_k=0, W^u_k(0)=1, \forall k \in \mathcal{K}$ \\
    }
    \KwResult{$A^u(t), \forall t=\{1,2,\cdots,T\}$}
    \For {$t=1$ \textnormal{to} $T$}
    {
    	\eIf{$\min_{k \in \mathcal{K}}\{N^u_k\} \leq l^\textnormal{EXP}$}
        	{Select action $k \in \mathcal{K}$ by round-robin selection}
        	{Select action $k \in \mathcal{K}$ by PDF selection}
    	Set $A^u(t)=I_k$ and perform the action \\
    	$R^u_k(t)=
    	\begin{cases} 
        	1 \quad & \textnormal{if ACK is received} \\
        	0 \quad & \textnormal{otherwise}.
    	\end{cases}$ \\
    	$P^u_k(t+1) = (1-\gamma) \Big( \frac{W^u\_k(t)}{\sum\_{k \in \mathcal{K}}{W^u\_k(t)}} \Big) + \frac{\gamma}{K}$ \\
    	$P^u_k(t+1) = \frac{P^u_k(t+1)}{\sum_{k \in \mathcal{K}}{P^u_k(t)}}$ \\
    	$W^u_k(t+1) = W^u_k(t) \times exp\Big(\frac{\gamma R^u_k(t)}{K \times P^u_k(t+1)} \Big)$ \\
    	$N^u_k = N^u_k +1$ \\ 
    	\If{$N^u_k > l^\textnormal{EXP}$ \&\& $P^u_k(t+1) < \frac{1}{2} \max_{\forall k \in \mathcal{K}}\{P^u_k(t)\}$}{
            {$P^u_k(t+1) = 0$} \\
    	}
    	\If{${N^u_k > \alpha_u \times l^\textnormal{EE}}$}{
        	{${N^u_k=0, \forall k \in \mathcal{K}}$ \\
        	${\alpha_u=\alpha_u+1}$}
    	}
    }
\end{algorithm} 
% \setlength{\textfloatsep}{0pt}
% \vspace{-5mm}
\subsection{The pre-processing phase of MIX-MAB algorithm}
Let $N^u_k$ represent the number of times that $u$th ED selects the $k$th action. At the first time, when the $u$th ED has a packet to send, it selects the first configuration setting, i.e., $I_0$. The next packet of the $u$th ED takes the action of $I_1$ of setting parameters in a round-robin manner and so on (line 8 of Algorithm \ref{al}). This process, taken from the SE algorithm, continues until the $u$th ED takes all actions once. Then, the exploration process is repeated for $l^\textnormal{EXP}$ times starting from the beginning by selecting $I_0$ to $I_{K-1}$ (line 7). \par Based on our simulation results, setting $l^\textnormal{EXP}=5$ leads to a short exploration, generating the best results. Each time $u$th ED takes action, the weight and probability of that action are updated based on the reception of the ACK message. Lines 14-16 in Algorithm \ref{al} present the equations for calculating the probability and weight of an especial index for the following iteration. According to these equations, taken from the EXP3 algorithm, if $k$th action does not result in receiving the ACK message by the $u$th ED, i.e., $R^u_k(t)=0$, the relevant weight and probability of that action, i.e., $W^u_k$ and $P^u_k$ are not updated, so they remain as the value in the previous iteration. Note that the summation of all probabilities should be one. So, after calculating the probability using line 14, we normalize the probabilities in line 15.~In~Fig.~\ref{fig:rl},~\cmark~and~\xmark~represent the reception and non-reception of ACK, respectively. As seen in Fig. \ref{fig:rl}(a), the value of $N^u_k$ for all $k \in \mathcal{K}$ reaches $l^\textnormal{EXP}$ at the end of the pre-processing phase. The whole actions' weights and probabilities are the first phase's output, which will be used as input in the second phase.
\subsection{The main processing phase of MIX-MAB algorithm}
After finishing the first phase, the second phase, i.e., the main process, starts as shown in Fig. \ref{fig:rl}(b). In this phase, the actions are selected based on their probability density function (PDF) values (line 10). Based on this condition, which originated from the EXP3 scheme, actions with a high probability have a higher chance of being chosen. Taking individual action $k$ in each iteration $t$, the $u$th ED updates that action's weight and probability by receiving the ACK. Suppose the updated probability of the $k$th action is smaller than half of the maximum probability of all actions in $\mathcal{K}$. In that case, the $u$th ED removes the $k$th action from its available actions not to be selected in subsequent steps (lines 18-20). This removal process is done implicitly by setting the probability of the $k$th action equal to zero. Note that the above threshold for removing an action is obtained heuristically through the simulation results, leading to the best performance. This process continues until the number of times one of the actions is selected, e.g., $N^u_k$ reaches a threshold value (lines 21-24). Actually, the action with the greatest $N^u_k$ is the most suitable one with the best transmission parameter configuration. One of our novelties in this phase is considering dynamic values for the threshold. Let $\alpha_u \times l^\textnormal{EE}$ denote the considered threshold, where $l^\textnormal{EE}=100$ is a constant parameter and $\alpha_u$ is a variable parameter. At the initialization, we set $\alpha_u=1$ making the threshold be equal to $l^\textnormal{EE}=100$. When $N^u_k$ for special action like $k$ reaches the threshold of $100$, the $\alpha_u$'s value is increased by one leading to a grown threshold value to $2 \times l^\textnormal{EE}=200$ and so on. Whenever the value of $N^u_k$ of one action achieves the threshold, the $N^u_k$ for all actions are set to zero while the weight and probabilities values remain unchanged. Through resetting the number of selected actions, the algorithm continues from line 7. So it starts the exploration process, i.e., pre-processing phase, again and gives a new chance to the actions that already have been removed in the second phase and may have become an appropriate option by changing the environmental conditions. Note that by not zeroing the weight and probabilities of the actions in the reset process, the proposed algorithm does not eliminate the previous experiences learned from the environment.
% \vspace{-5mm}
\section{Numerical Results}
This section evaluates our algorithm performance by simulation results and compares it with other approaches.
% \vspace{-5mm}
\subsection{Simulation Setup}
We use the LoRa-MAB simulator proposed in \cite{Ta} and customize it to evaluate our algorithm performance\footnote{This framework is available in https://github.com/Farzad-Azizi/MIX-MAB}. We consider a LoRaWAN network composed of one GW, located at the center of a disc-shaped cell of radius $r=4.5$ km, with $N=100$ EDs uniformly distributed as in \cite{Ta}. We use the log-distance path loss model, presented in \cite{Bor}. The $1\%$ DC limitation is satisfied by setting the packet generation rate of each ED to $15$ packets/hour and the packet length of $50$ bytes generated through an exponential distribution. We assume that there is no collision between the ACK and uplink messages. The GW delivers the ACK message on a separate channel with a higher DC. Thus, when the ED does not receive the ACK, it shows that the packet has been lost. We also consider the inter-SF collision and capture effect. We run the simulations for $72 \times 10^{10}$ millisecond, which is equal to $200$ KHours horizon time. Other parameters that affect the performance of LoRaWAN, including bandwidth and coding rate, are set as $125$KHz and $4/5$, respectively. \par To evaluate the proposed algorithm performance and compare it with other schemes, we use three following metrics:
\begin{enumerate}
    \item \textbf{Convergence time}: an RL algorithm converges when the learning curve gets flat and no longer increases.
    \item \textbf{PDR}: described the total received packets by the GW divided by the total sent packet from all EDs.
    \item \textbf{EC}: defined as the average EC per transmitted packet per ED.
\end{enumerate}
We also consider five following scenarios:
\begin{itemize}
    \item \textit{\textbf{Scenario 1:}} Each ED can select one SF $\in [7,12]$, SC and TP are fixed at SC $=868100$ Hz and TP $=14$ dBm.
    \item \textit{\textbf{Scenario 2:}} Each ED can select one SF $\in [7,12]$, one SC $\in \{868100, 868300, 868500\}$ Hz, and TP is fixed at TP $=14$ dBm.
    \item \textit{\textbf{Scenario 3:}} Each ED can select one SF $\in [7,12]$, one TP $\in \{8, 11, 14\}$ dBm, and SC is fixed at SC $868100$ Hz.
    \item \textit{\textbf{Scenario 4:}} Each ED can select one SF $\in [7,12]$, one SC $\in \{868100, 868300, 868500\}$ Hz, and one TP $\in \{8, 11, 14\}$ dBm.
    \item \textit{\textbf{Scenario~5:}}~Options~are~like~the~previous~scenario, but the number of packets sent by EDs changed from~1~packet per hour to~1~packet per day and~1~packet~per~week.
\end{itemize} 
% \vspace{-5mm}
\begin{figure}[b]
    \centering
    \includegraphics[scale=0.25]{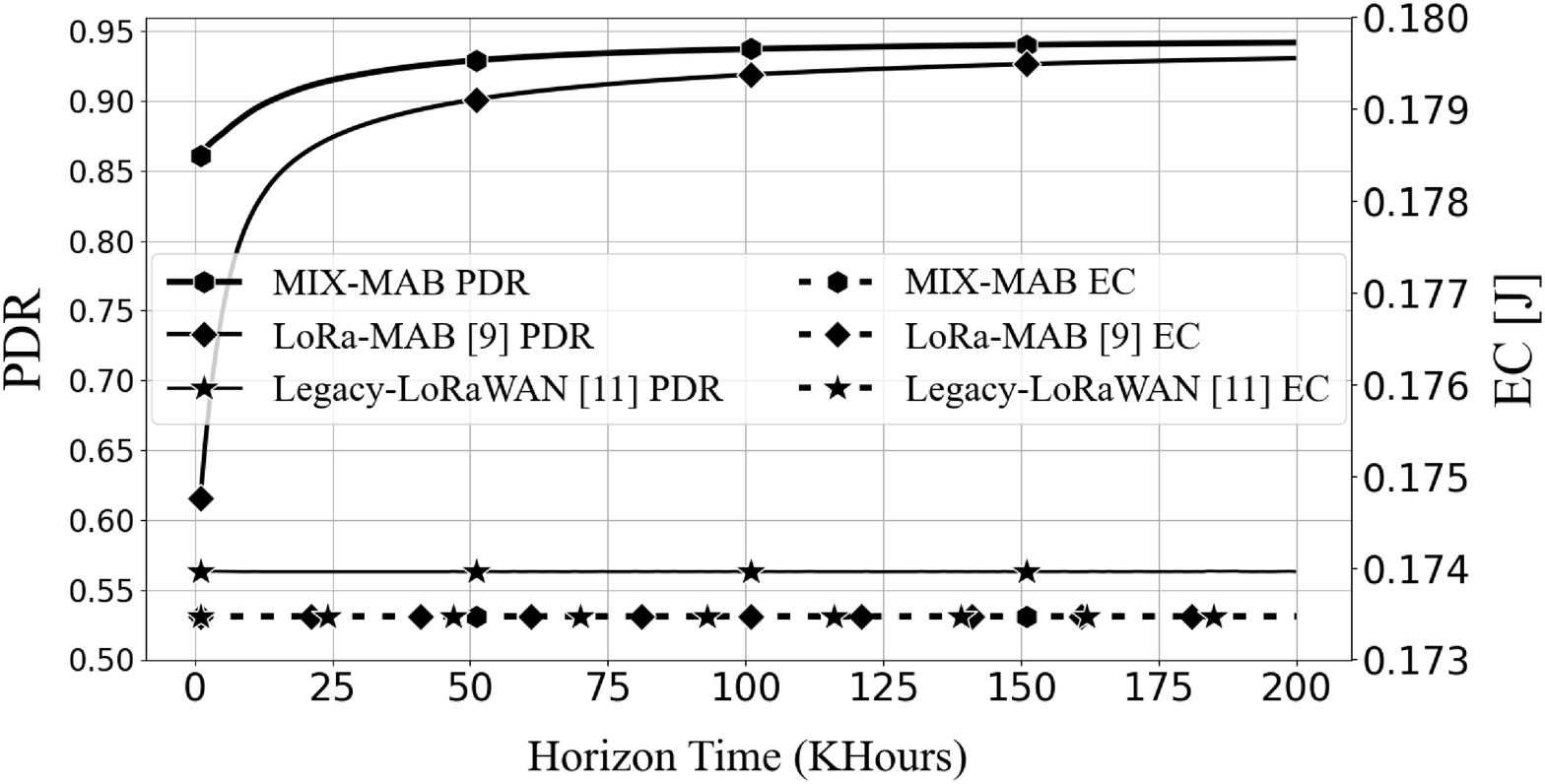}
    \vspace{-8mm}
    \caption{PDR \& EC in MIX-MAB, LoRa-MAB, and LoRa in Scenario 1.}
    \label{fig:e1}
\end{figure} 
% \vspace{-5mm}
\setlength{\textfloatsep}{0pt}
% \vspace{-5mm}
\subsection{Simulation Results}
Now, we provide the simulation results comparing the performance of our proposed algorithm with the LoRa-MAB \cite{Ta} and the Legacy LoRa \cite{Bor} algorithms in five defined scenarios. The LoRa-MAB algorithm proposed in \cite{Ta} is based on RL that uses the EXP3 scheme. The EDs configure their transmission parameters randomly in the Legacy LoRa method.
\subsubsection{Scenario 1} 
Fig. \ref{fig:e1} shows the PDR and EC in MIX-MAB, LoRa-MAB, and Legacy LoRa in Scenario 1. As we observe, the PDR of our proposed solution is higher than LoRa-MAB and Legacy LoRa algorithms. This is due to the nature of our proposed algorithm, which uses a combination of short-term exploration at the first phase and long-term exploitation and exploration processes at the second phase together. On the contrary, LoRa-MAB does not apply the short-term exploration initially, and Legacy LoRa does not employ the learning process at all. Additional to that, the probability initialization in LoRa-MAB is defined as a uniform distribution, i.e., $P^u_k(0)=\frac{1}{K},\forall k\in\mathcal{K},u\in\mathcal{U}$. However, we do not assume equal probability initialization for each action and ED in MIX-MAB. Instead, we set the initial probabilities as an undefined number, i.e., $nan$, and then update them based on the ACK reception. Besides that, MIX-MAB gives a second chance to all removed actions, but LoRa-MAB does not. \par We also see in Fig. \ref{fig:e1} that the EC of the MIX-MAB is the same as the Legacy LoRa and LoRa-MAB. The reason is that each ED transmits with TP of 14 $dBm$ in all algorithms leading to the same EC. Another important observation from Fig. \ref{fig:e1} is that our proposed algorithm's PDR converges after $100$ KHours in horizon time while the LoRa-MAB converges after $200$ KHours. Therefore, the convergence time of the proposed solution is half of the LoRa-MAB. Actually, uniform probability initialization in LoRa-MAB causes wrong choices to take a long time removing from the actions, leading to an increase in the convergence time. The Legacy LoRa scheme is not a learning-based algorithm, so its PDR and EC remained unchanged in time. 
% \vspace{-5mm}
\begin{figure}[b]
    \centering
    \includegraphics[scale=0.25]{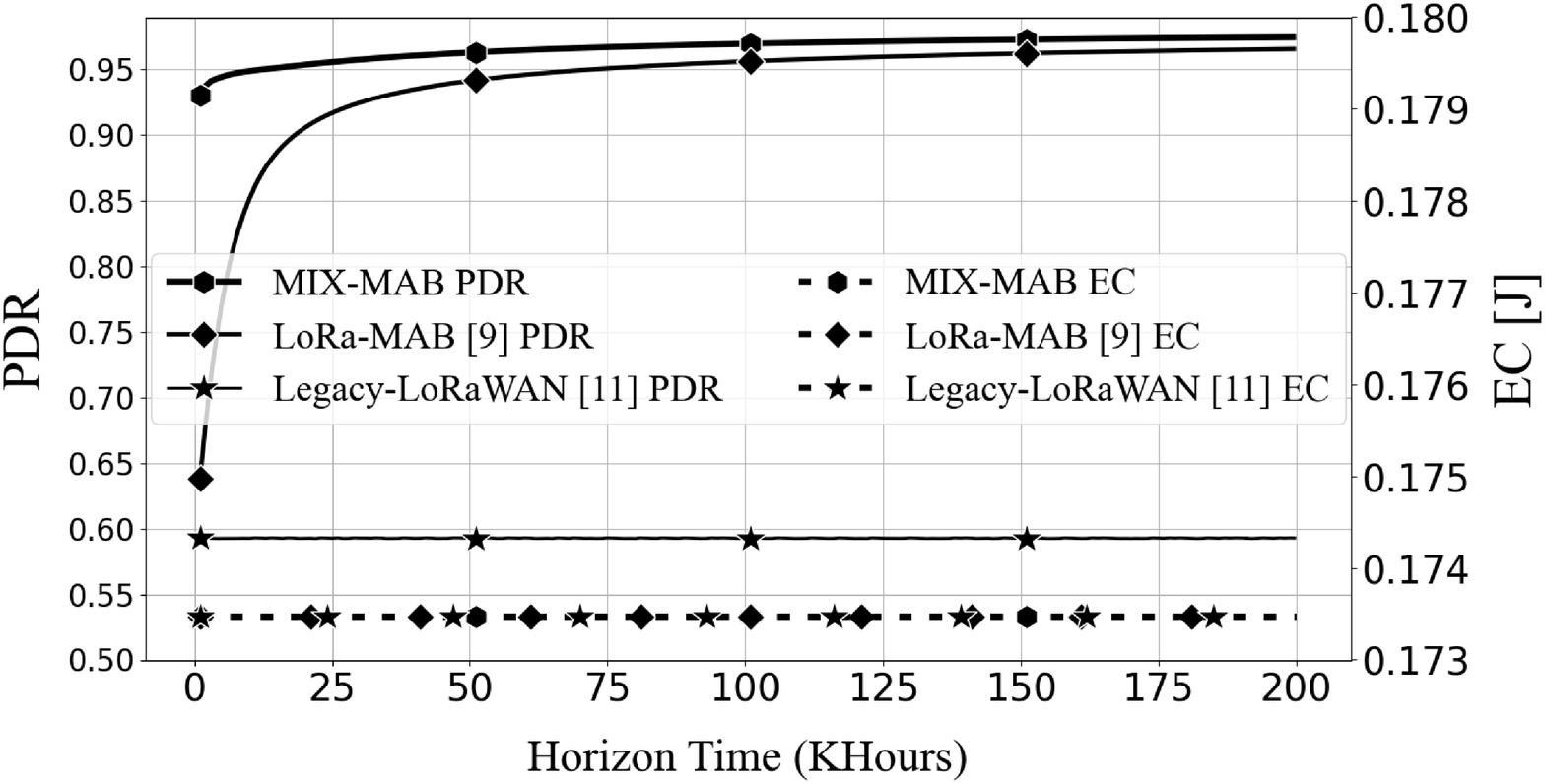}
    \vspace{-8mm}
    \caption{PDR \& EC in MIX-MAB, LoRa-MAB, and LoRa in Scenario 2.}
    \label{fig:e2}
\end{figure} 
% \vspace{-5mm}
\setlength{\textfloatsep}{0pt}
\subsubsection{Scenario 2} 
In this scenario, we evaluate the effect of SCs selection on overall performance. For this purpose, the EDs have the freedom of choosing one of three SCs compared to Scenario 1. Fig. \ref{fig:e2} shows the PDR and EC in MIX-MAB, LoRa-MAB, and Legacy LoRa in Scenario 2. As seen, the PDR of the MIX-MAB is higher than that in scenario 1. Thus, we can conclude that increasing the number of SCs increases the PDR. We also see that in Scenario 2, the PDR of the MIX-MAB is still higher than the Legacy LoRa and LoRa-MAB while their consumed power is the same.
\subsubsection{Scenario 3} 
\begin{figure}[t]
    \centering
    \includegraphics[scale=0.25]{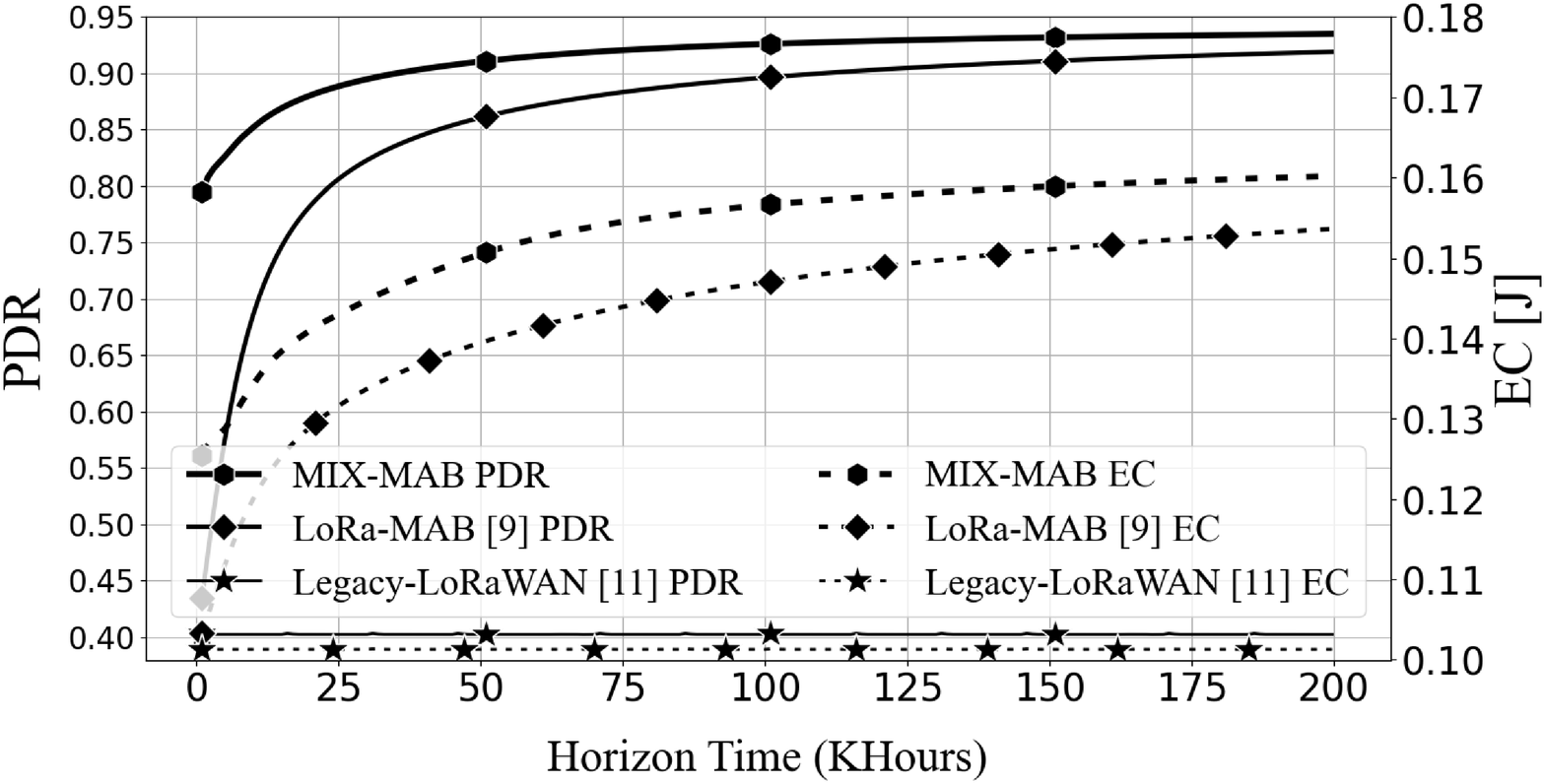}
    \vspace{-8mm}
    \caption{PDR \& EC in MIX-MAB, LoRa-MAB, and LoRa in Scenario 3.}
    \label{fig:e3}
\end{figure}

In this scenario, we investigate the effect of TP on the proposed algorithm performance. For this goal, EDs have the freedom of transmitting with one of three TP values compared to Scenario 1. Fig. \ref{fig:e3} shows the PDR and EC in MIX-MAB, LoRa-MAB, and Legacy LoRa in Scenario 3. As observed, the EC of the proposed algorithm is lower than that in Scenario 1, while its PDR is decreased. The reason is that having different values of TP to select by the EDs results in reduced TP, leading to lower EC. However, sending packets with less TP reduces PDR compared to Scenario 1, where all packets are sent with maximum TP. From this figure, we also see that the PDR of the MIX-MAB is still higher than the Legacy LoRa and LoRa-MAB. However, the EC of the MIX-MAB is a little bit more than Legacy LoRa and LoRa-MAB. This is due to the goal of MIX-MAB, to achieve the maximum PDR; thus, the ED transmits data with higher TP leading to higher EC.
% \vspace{-5mm}
% \vspace{-5mm}
\setlength{\textfloatsep}{0pt}
\subsubsection{Scenario 4} 
\begin{figure}[t]
    \centering
    \includegraphics[scale=0.25]{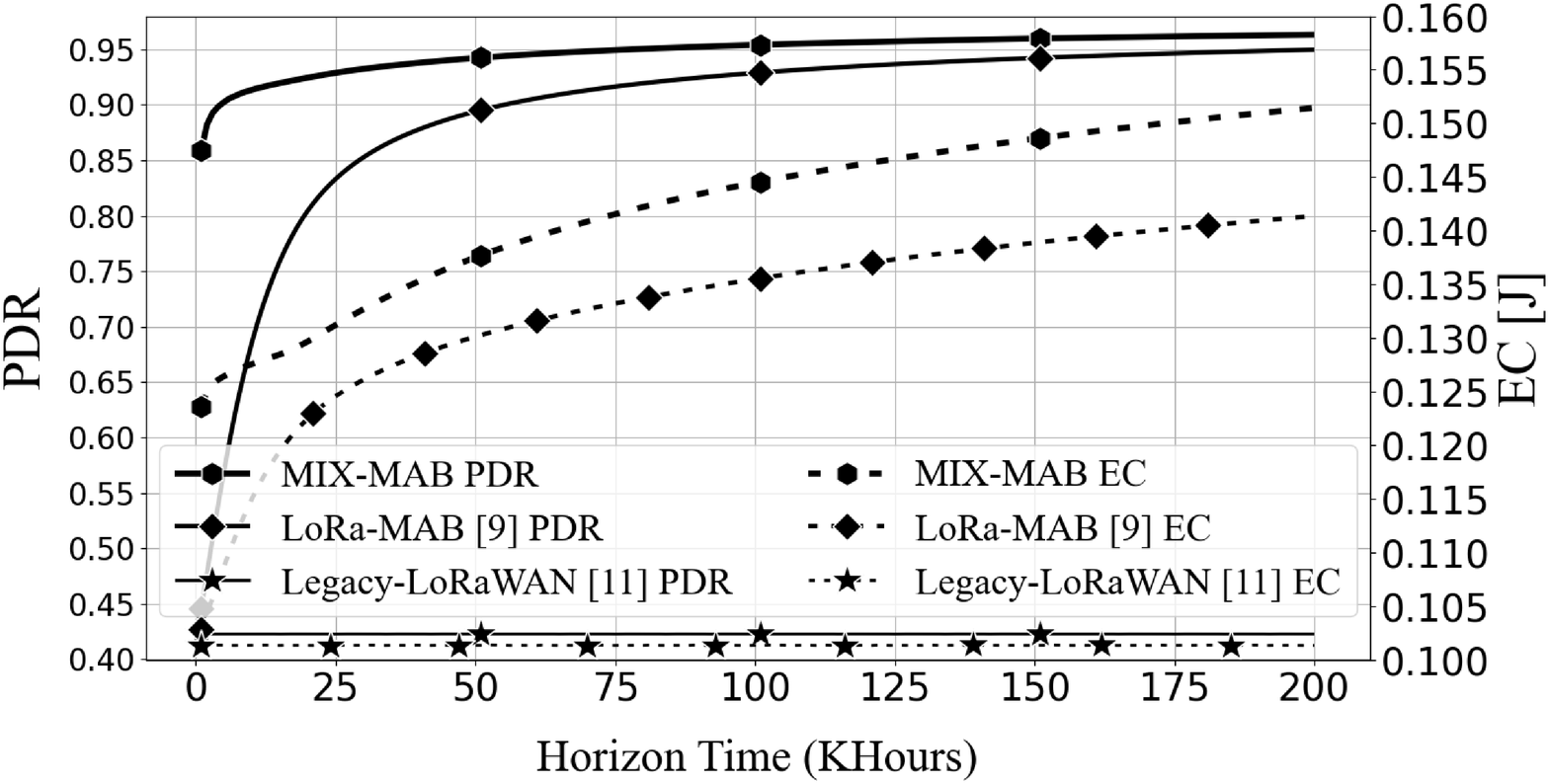}
    \vspace{-8mm}
    \caption{PDR \& EC in MIX-MAB, LoRa-MAB, and LoRa in Scenario 4.}
    \label{fig:e4}
\end{figure}

Here we evaluate the MIX-MAB performance when the EDs have the most freedom to select transmission parameters. Fig. \ref{fig:e4} shows PDR and EC in MIX-MAB, LoRa-MAB, and Legacy LoRa in Scenario 4. As shown, the PDR of the MIX-MAB is higher than that in Scenario 1, while the EC decreased. It concludes that the proposed algorithm achieves a high PDR when the EDs configure all three parameters. At the same time, it consumes lower energy than the other three previous scenarios. We also see that, similar to Scenario 2, the PDR of MIX-MAB is higher than the Legacy LoRa and LoRa-MAB at the cost of higher EC.
% \vspace{-5mm}
% \vspace{-5mm}
\setlength{\textfloatsep}{0pt}
\subsubsection{Scenario 5} 
\begin{figure}[t]
    \centering
    \includegraphics[scale=0.25]{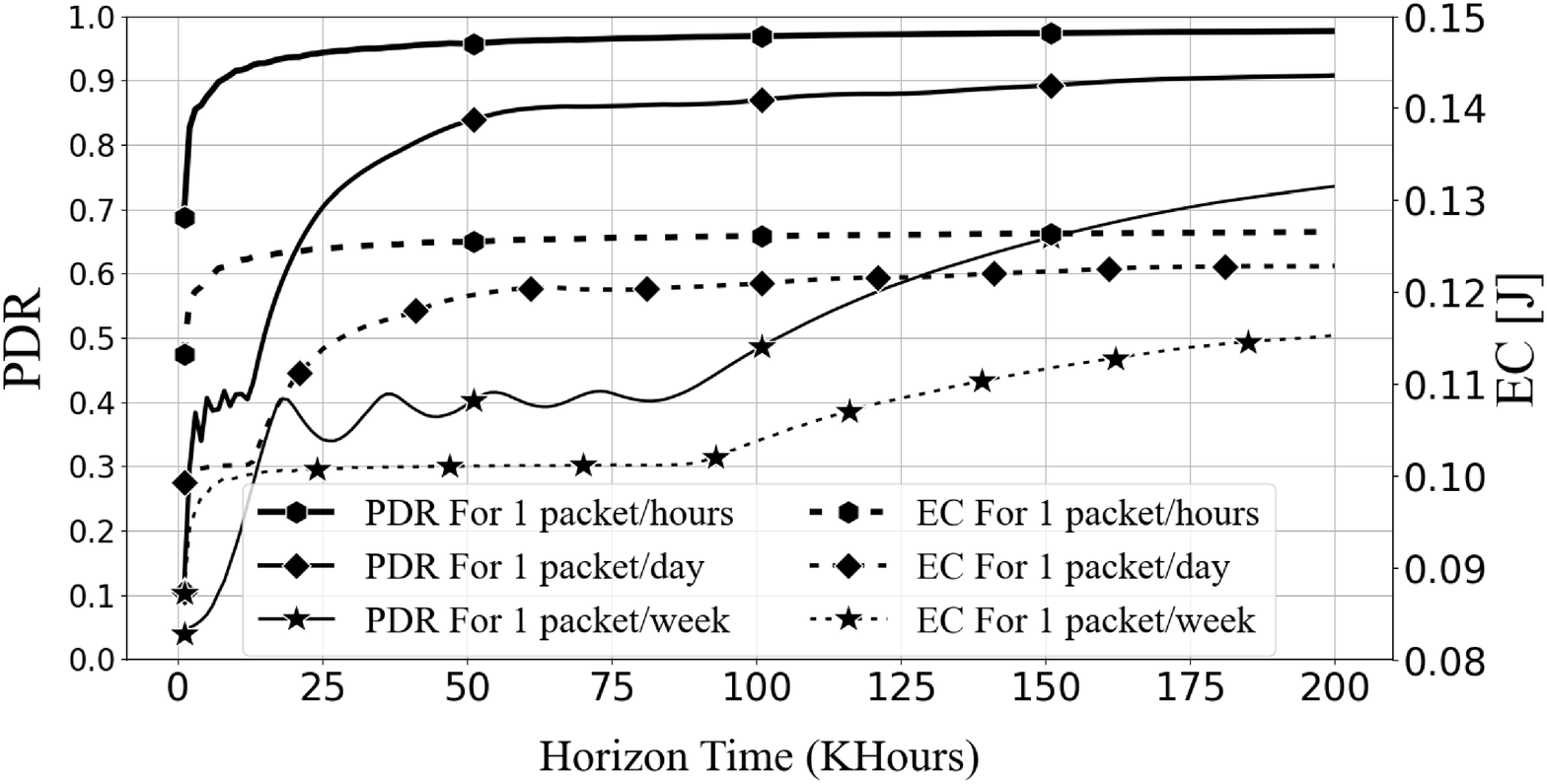}
    \vspace{-8mm}
    \caption{number of transmitted packets’ effects on EC \& PDR.}
    \label{fig:e5}
\end{figure}

This scenario evaluates the number of transmitted packets' effects on EC and PDR of the proposed algorithm. The setting is similar to Scenario 4, except that packet transmission frequency varies. As seen in Fig. \ref{fig:e5}, increasing the frequency of sent packets leads to both improving the EC and the PDR due to receiving more feedback from the NS, which in turn leads to an improvement in the learning process. So, based on observations and considering the trade-off between results, we have the best outcome for 1 packet per day, which is more consistent with the IoT application and fulfills its requirements.
% \vspace{-3mm}
\section{Conclusion and Future Works}
This paper focused on improving LoRaWAN's resource allocation algorithm in terms of PDR. We presented an RL-based resource allocation algorithm enabling the LoRa EDs to configure their transmission parameters in a distributed manner. Our proposed MIX-MAB algorithm combines two MAB schemes, i.e., SE and EXP3, making the LoRa more efficient in interference management, leading to higher network throughput. We evaluated the performance of the MIX-MAB and compared it with the Legacy LoRa and LoRa-MAB mechanisms in five different scenarios. Our simulation results show that the convergence time of the MIX-MAB is half of LoRa-MAB while our solution achieves higher PDR than Legacy LoRa and LoRa-MAB in all scenarios. We will reduce the computational overhead for future works by proposing a new version of the MIX-MAB algorithm on the NS side. Furthermore, we will deploy and evaluate our proposed algorithm for QoS requirements of various industrial applications, including unmanned aerial vehicles.
% \vspace{-5mm}
% \vspace{-5mm}
\setlength{\textfloatsep}{0pt}

\end{document}